\documentclass[aps,prb,twocolumn,groupedaddress,showpacs]{revtex4-1}
\usepackage{epsfig}
\usepackage[dvipsnames,usenames]{color}
\usepackage[normalem]{ulem}
\usepackage{amsmath}
\usepackage{xfrac}
\emergencystretch=\maxdimen
\begin{document}

\title{Magnetic Order-Disorder Transitions on a 1/3 - Depleted
Square Lattice}

\author{H.-M.~Guo$^{1,3}$,
T. Mendes-Santos$^{2,3}$,
W.E.~Pickett$^{3}$,
and R.T.~Scalettar$^{3}$
}
\affiliation{$^1$Department of Physics,
Beihang University, Beijing, China}
\affiliation{$^2$Instituto de Fisica, Universidade Federal do Rio de
Janeiro Cx.P. 68.528, 21941-972 Rio de Janeiro RJ, Brazil}
\affiliation{$^3$Physics Department, University of California, Davis,
California 95616, USA}

\begin{abstract}
Quantum Monte Carlo simulations are used to study the magnetic and
transport properties of the Hubbard Model, and its strong coupling
Heisenberg limit, on a one-third depleted square lattice.  This is the
geometry occupied, after charge ordering, by the spin-$\frac{1}{2}$
Ni$^{1+}$ atoms in a single layer of the nickelate materials La$_4$Ni$_3$O$_8$ and (predicted) La$_3$Ni$_2$O$_6$.  Our model is also a
description of strained graphene, where a honeycomb lattice has bond
strengths which are inequivalent.  For the Heisenberg case, we determine
the location of the quantum critical point (QCP) where there is an onset
of long range antiferromagnetic order (LRAFO), and the magnitude of the
order parameter, and then compare with results of spin wave theory.  An
ordered phase also exists when electrons are itinerant.  In this case,
the growth in the antiferromagnetic structure factor
coincides with the
transition from band insulator to metal in the absence of interactions.
\end{abstract}

\pacs{
71.10.Fd, 
75.47.Lx, 
}

\maketitle
\noindent

\section{Introduction:}

Over the last several decades, quantum Monte Carlo (QMC) methods have
been widely used to investigate magnetic, charge, and pairing
correlations in the Hubbard Hamiltonian on a square lattice
\cite{white89a,scalapino94,vilk97,zhang97b,maier00,capone07,gull07}.  A
central issue has been the intimate interplay between these
different types of order, most fundamentally the possibility that
magnetic correlations give rise to $d$-wave superconductivity.  The
occurrence of inhomogeneous (stripe) charge distributions upon doping the
half-filled lattice, where antiferromagnetism (AF) survives in regions of low hole
concentration but is suppressed on stripes of high concentration, has
also been shown to have profound implications for
pairing\cite{berg09}.

In more recent studies, the effect of depletion of the square lattice has also
been investigated.  In this case, a regular removal of sites can be
regarded as an extreme limit of the spontaneous formation of charge and
spin patterns in which the degrees of freedom on certain sites are
completely eliminated.  Further types of transitions were then shown to
occur within these geometries.  Two prominent examples are the Lieb
lattice\cite{lieb89}, where 1/4 of the sites are removed, giving rise to
a flat electronic band and ferromagnetism, and the 1/5 depleted lattice
\cite{n_katoh_95,k_ueda_96,m_troyer_96,m_gelfand_96,cavo_wep} where spin
liquid phases compete with magnetic order.  This latter geometry is
realized by the vanadium atom locations in CaV$_4$O$_9$, and also by
some members of the iron-pnictide family\cite{w_bao_11,f_ye_11}.  A
crucial feature of this situation is the occurrence of two separate
types of bonds, and hence of exchange or hopping energies, in the
depleted structure.

Depleted lattices can also be formed starting from other, non-square,
lattices.  For example, the Kagom\'e lattice arises from removing one
fourth of the sites of a triangular lattice.  Like the Lieb lattice, the
Kagom\'e structure has a flat band.  However, because it is not bipartite,
the band does not lie between the dispersing ones.

\begin{figure}[!h]
\includegraphics[width=5cm]{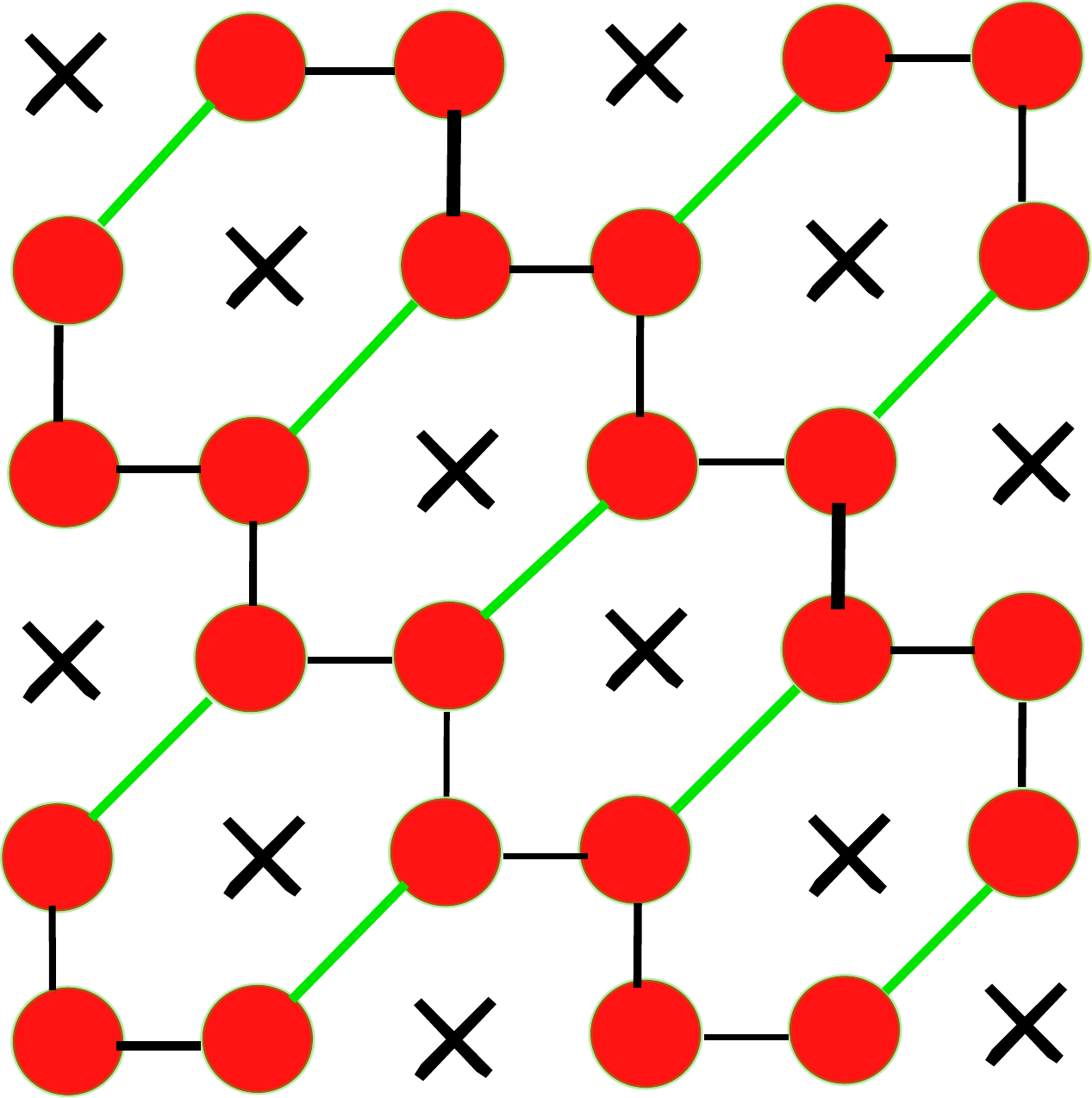}
\caption{The one third depleted square lattice.  A regular
diagonal stripe array
of black crosses is removed, leaving the red site structure.  We will
assume two types of bonds exist corresponding to connections between NN
(black)
and NNN (green) sites of the original square geometry.  (See text.)
\label{fig1}}
   \end{figure}

In this paper, we investigate the magnetic and charge patterns within
the 1/3 depleted square lattice of Fig.~\ref{fig1}, which is formed by
the red sites remaining after the removal of the black sites,
which form stripes along one diagonal.
The bonds between red sites are of two sorts:  ones which were the near
neighbor bonds of the original, full square lattice, and ones which
connect through the diagonal rows of removed sites, and which were next
near neighbors of the original lattice.  This distinction will be
modeled, in the following sections, by allowing for different energy
scales on the two types of bonds.
Notice that this lattice structure remains bipartite, a fact
which has implications for AF order without frustration and
also for the absence of a sign problem in QMC simulations.

Figure \ref{fig1} is equivalent to a strained version of the honeycomb geometry realized in
graphene.  ``Artificial graphene" lattices, can be achieved by
nanopatterning\cite{park09}, by molecule-by-molecule
assembly\cite{gomes12}, or by trapping ultracold atoms on optical
lattices.  They offer the possibility of tunable bond strengths, for
example through application of strain, and have recently been discussed
as a means for further investigation of Dirac particles and their
associated correlated and topological phases\cite{polini13}.
Graphene with a ``Kekul\'e distortion" \cite{hou07,roy10,gomes12}.
involves the appearance of two distinct
bond hoppings,
albeit in a pattern different from that of Fig.~\ref{fig1}.

A second motivation for investigating the
geometry of Fig.~\ref{fig1}, which more
directly connects with the notion of `depletion' and which
also fundamentally involves magnetic order, is provided
by recent experimental\cite{zhang16} and
theoretical\cite{botana16} studies of the layered nickelates
La$_4$Ni$_3$O$_8$,
and La$_3$Ni$_2$O$_6$.
In these materials, the formal Ni valences
of +1.33 and +1.5 are separated into charge ordered Ni$^{1+}$
(spin $\frac{1}{2}$) and Ni$^{2+}$ (spin 0), so that spin-$\frac{1}{2}$
stripes are formed at 45$^{\circ}$ relative to
the Ni-O bonds, as in Fig.~\ref{fig1}
for La$_4$Ni$_3$O$_8$.
This charge ordering is accompanied by structural distortions
and the opening of a gap.
The Ni$^{1+}$ atoms form an AF arrangement in analogy with
the magnetism of the CuO$_2$ planes of the cuprate superconductors.
Here we will investigate AF correlations associated with this
geometry.  Other layered nickelate
materials\cite{tranquada94,sachan95,yoshizawa00,kajimoto03}
have also been investigated
with quantum simulations, especially within the
classical spin-fermion method\cite{hotta04}.

   \begin{figure}[t]
   \includegraphics[width=7.0cm]{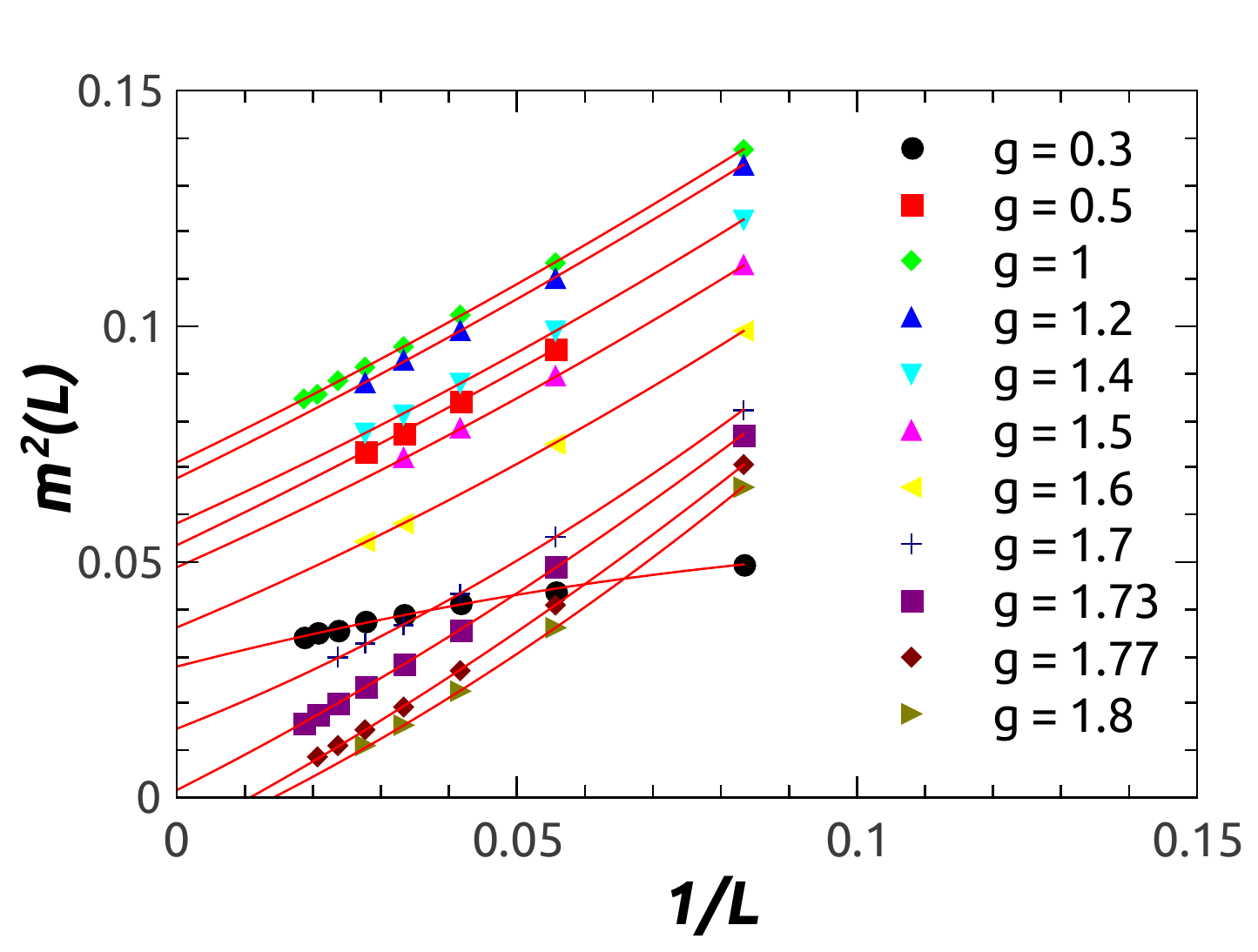}
   \caption{Finite size scaling of the square of the AF order parameter
$m^2$ for the spin-1/2 Heisenberg model.
For a ratio $g > 1.75 = g_c$ of exchange couplings, a transition to a disordered
spin liquid state occurs.  LRAFO persists to small values of the
interchain exchange.  Data were obtained with the SSE algorithm.
\label{fig2}}
   \end{figure}

\section{Strong Coupling (Heisenberg) Limit}

We first consider the case of localized spin-1/2 moments
on the 1/3 depleted lattice with Hamiltonian
\begin{align}
H = J \, \big[\,  \sum_{\langle i j \rangle} \vec S_{i} \cdot \vec S_{j}
+g \sum_{\langle \langle i j \rangle \rangle}
\vec S_{i} \cdot \vec S_{j} \, \big]
\label{eq:hamiltonian}
\end{align}
with exchange constants $J$ and $gJ$ on the two types of bonds of
Fig.~\ref{fig1}.

This model can be treated within linear spin wave theory (LSWT)
by replacing the spin
operators by bosonic ones via the Holstein-Primakoff (HP) transformation, and
then invoking the linear approximation describing small fluctuations
around the broken symmetry phase.  The resulting noninteracting
Hamiltonian can be diagonalized in momentum space and
through a Bogliubov rotation.  The spin wave spectrum is,
\begin{align}
\omega(J^*,k) &= J^*\sqrt{1 - \frac{|\gamma(\vec k)|^2}{J^{*2} } },
\end{align}
where,
\begin{align}
\gamma(\vec k) &= \sum_\delta J(\delta) e^{-i \vec k \cdot \vec r_\delta}
\nonumber \\
&=J \big[ e^{-i ((\vec k \cdot \vec a_1) + (\vec k \cdot \vec a_2))/3} +
e^{i ((\vec k \cdot \vec a_1)- 2 (\vec k \cdot \vec a_2))/3} \big]
\nonumber\\
&+gJ e^{i (2(\vec k \cdot \vec a_1)- (\vec k \cdot \vec a_2))/3}
\label{lswtdisp}
\end{align}
with lattice vectors $\vec a_1 = 2\hat x - \hat y$ and
$\vec a_2 = \hat x + \hat y$.
Here $J^* = \sum_\delta J(\delta)$ is the sum
of exchange constants over near neighbor sites.
The AFM staggered order parameter,
\begin{align}
m_s= \frac{1}{N} \Big( \sum_{i \in A} \langle S_i^z \rangle - \sum_{i \in B} \langle S_i^z \rangle \Big).
\label{lswmag}
\end{align}
is obtained in the LSWT, writing $\langle S_i^z \rangle$ in terms of HP
operators.  At $T=0$, we obtain:

\begin{align}
m_s=S+\frac{1}{2}
-\frac{1}{N} \sum_{\vec k}
 \Big( 1 - \frac{|\gamma(\vec k)|^2}{J^{*\,2}} \Big),
\label{lswtop}
\end{align}
where $S$ is the spin.

    \begin{figure}[t]
   \includegraphics[width=7.0cm]{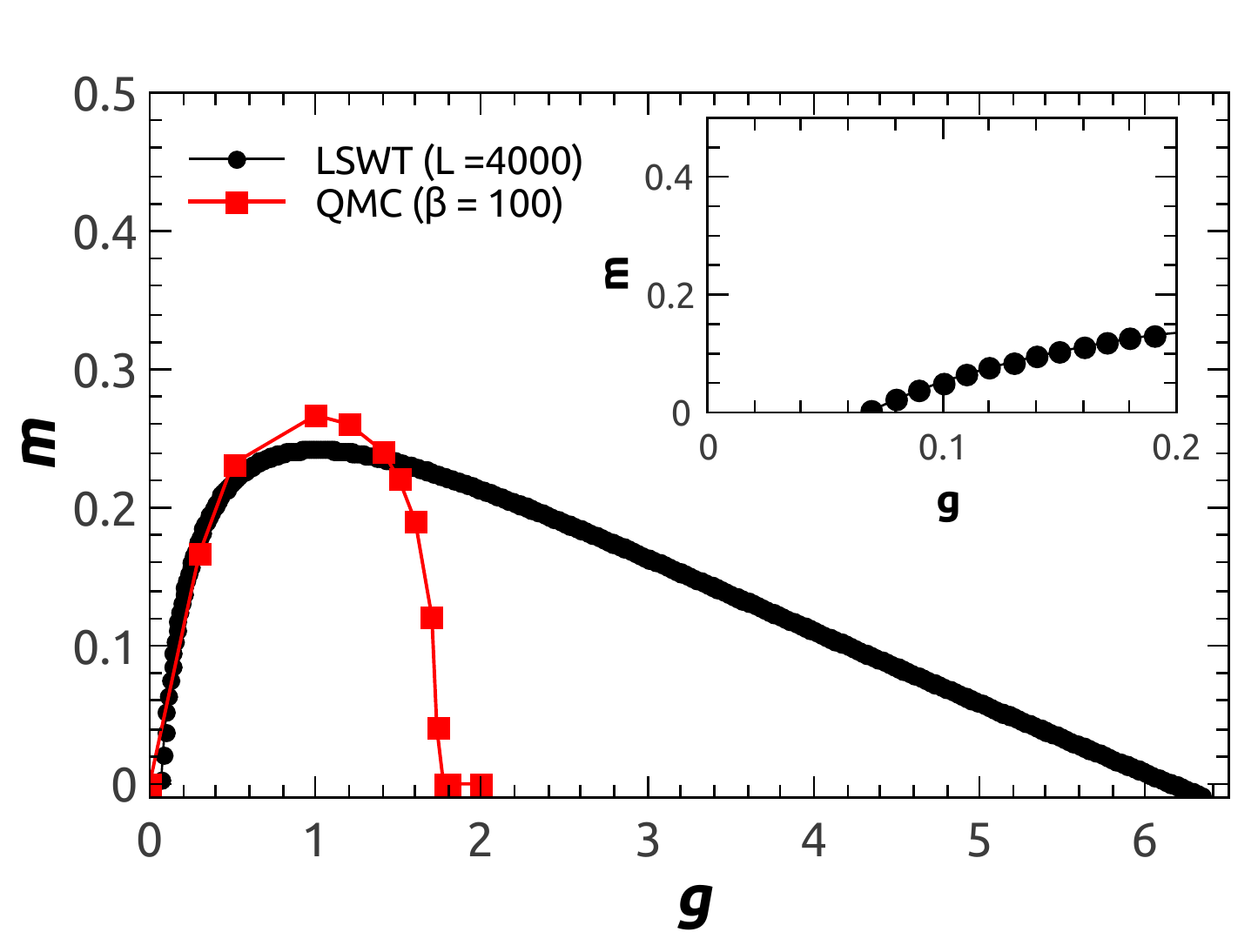}
   \caption{Extrapolated values of the SSE results for the AF order
parameter from Fig.~\ref{fig2} and the results of LSWT analysis,
Eq.~\ref{lswtop}.  With LSWT(SSE), LRAFO disappears above $g_c =
6.20 \pm 0.02 \,\, (1.75 \pm 0.01)$.
\label{fig3}}
   \end{figure}

We can also treat the Hamiltonian more exactly on lattices
of finite size using the stochastic series expansion (SSE)
quantum Monte Carlo method\cite{syljuasen02,sandvik94}.
SSE samples terms in the power
expansion of $e^{-\beta \hat H}$ in the partition function.
Operator loop (cluster) updates perform the sampling
efficiently\cite{syljuasen02}.
The square of the staggered magnetization,
$\left<m_s^2\right>$,
can be evaluated to high precision, and extrapolated to
the thermodynamic limit.

Figure \ref{fig2} shows the results of SSE simulations for different
values of the bond anisotropy $g$ and inverse linear system size $1/L$.
The order parameter first increases with $g$, reaching a maximum at
the honeycomb limit $g=1$, and finally begins to decrease.
LRAFO vanishes above $g_c = 1.75 \pm 0.01$.
The extrapolated order parameter from SSE (Fig.~\ref{fig2}) and from
LSWT (Eq.~\ref{lswtop}) is given in Fig.~\ref{fig3}.
LSWT greatly
overestimates the persistance of LRAFO at large $g$.  It also
predicts a quantum phase transition at small, but nonzero,
$g_c = 0.065 \pm 0.005$.
Similar to the case of a square lattice with anisotropic exchange \cite{Sakai89,affleck94,sandvik99}, a zero $g_c$ is expected here though a small nonzero value is obtained in our calculations due to finite size effect.

We emphasize the contrast of these results with those of the
Heisenberg model on
1/5-depleted lattice\cite{m_troyer_96}
appropriate to modeling CaV$_4$O$_9$ where
the lower $g_c = 0.60 \pm 0.05$.
The difference, as for the case of the anisotropic square lattice,
is that for the 1/5 depleted case the building blocks are
small clusters (either dimers or four site plaquettes) in both
the $g$ small and $g$ large limits.  In the present case,
two site clusters are formed for large $g$, but
the small $g$ limit still has extended 1-d structures.
These give rise to LRAFO even for small $g$.

   \begin{figure}[!h]
   \includegraphics[width=8.5cm]{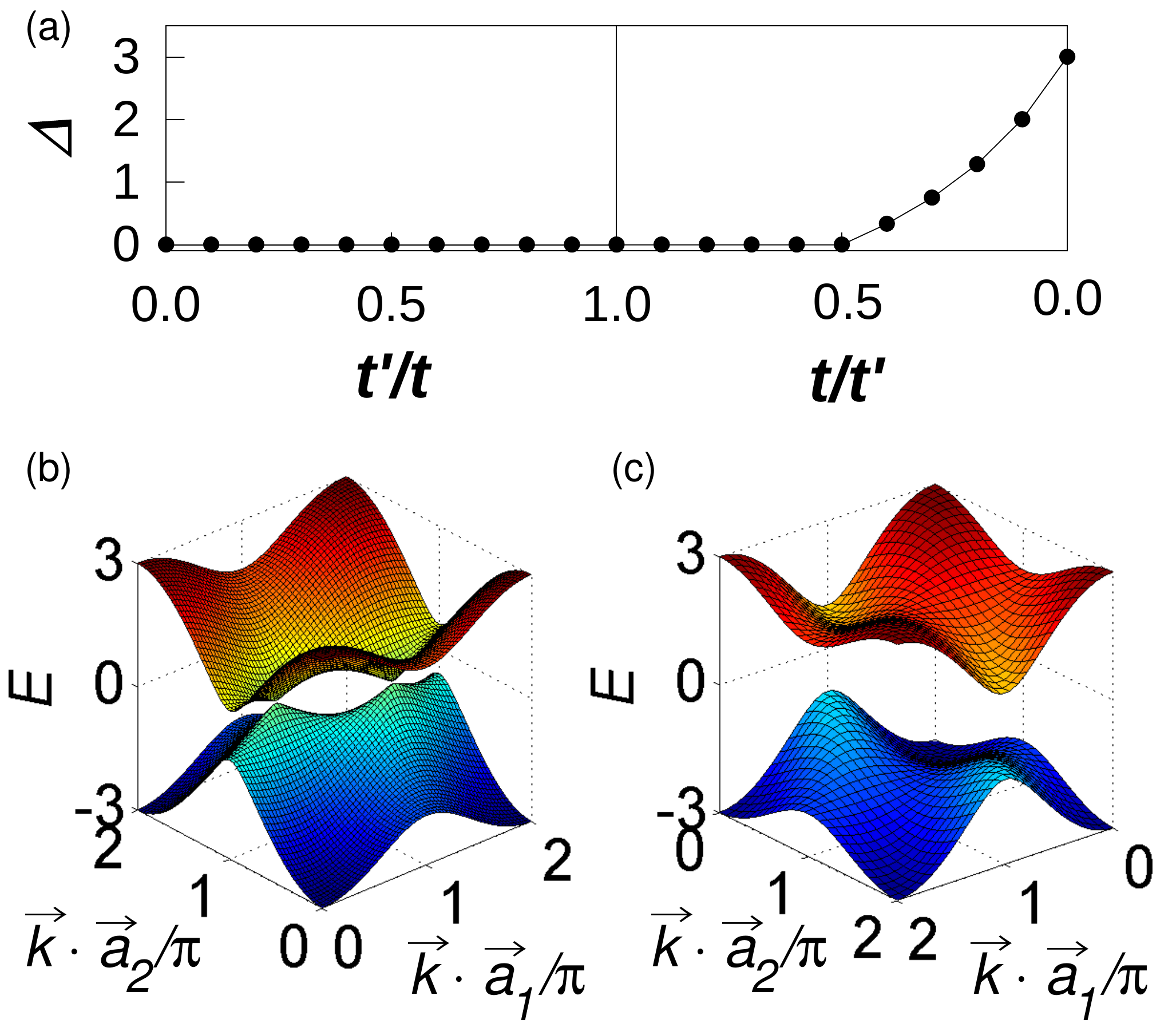}
   \caption{
(a) Band gap $\Delta$ as a function of the ratio
of hopping.  $\Delta$ vanishes for $t'/t<2$.  The noninteracting
limit is a band insulator ($\Delta > 0$) for $t'/t>2$.
(b)  Semi-metallic band stucture at
$t'/t=0.5$.
(c)  Insulating band stucture at
$t/t'=0.25$.
\label{fig4}}
   \end{figure}

\section{Itinerant Limit}

We next consider itinerant electrons, a single band Hubbard Hamiltonian
on the same 1/3-depleted lattice,
\begin{align}
H = -t  \, & \sum_{\langle i j \rangle\sigma}
\big( \, c^{\dagger}_{i\sigma} c^{\phantom{\dagger}}_{j\sigma}
+ c^{\dagger}_{j\sigma} c^{\phantom{\dagger}}_{i\sigma} \, \big)
\nonumber \\
-t' & \sum_{\langle\langle i j \rangle\rangle\sigma}
\big( \, c^{\dagger}_{i\sigma} c^{\phantom{\dagger}}_{j\sigma}
+ c^{\dagger}_{j\sigma} c^{\phantom{\dagger}}_{i\sigma} \, \big)
\nonumber \\
+U & \sum_i
\big( \, n_{i \uparrow} - \frac{1}{2} \, \big)
\big( \, n_{i \downarrow} - \frac{1}{2} \, \big)
\label{hubham}
\end{align}

The hoppings
along and between the
one dimensional chains are $t$ and $t'$, respectively.
The properties of this model are solved using the determinant QMC
method\cite{blankenbecler81}.  In this method the partition function
is expressed as a path integral.  The discretization of inverse
temperature $\beta$ enables the isolation of the quartic
interaction terms which are decoupled via a Hubbard-Stratonovich (HS)
transformation.  The resulting quadratic fermionic trace is
done analytically, and the HS field is then sampled
stochastically.  Because the scaling is cubic in
the lattice size $N$ we study systems only up to $N=2\times 12\times 12$
sites in contrast to the spin models described
in the previous section where SSE scales linearly in $N$
and systems up to $N = 1600$ (or more) are accessible.
Equation (\ref{hubham}) is written in particle-hole
symmetric form so that the lattice is half-filled
$\rho = \langle n_{i\uparrow}+n_{i\downarrow} \rangle = 1$
for all lattice sites $i$ and any values of $t',U$ and temperature $T$.
At this electron density, simulations are possible
down to low $T$ without encountering
the fermion sign problem\cite{loh90}.

In the noninteracting limit of
Eq.~(\ref{hubham}) we have two bands with dispersion,
\begin{align}
E(\vec k)=\pm
\Big[ \,\,
& \big(t+ t \, {\rm cos}(\vec k \cdot \vec a_2) +
t' \, {\rm cos}(\vec k \cdot \vec a_1) \big)^2
\nonumber \\
+ & \big(t \, {\rm sin}(\vec k \cdot \vec a_2) +
t' \, {\rm sin}(\vec k \cdot \vec a_1) \big)^2
\,\, \Big]^{1/2}
\end{align}
Here the noninteracting band width $w$ is kept fixed, $w=4t+2t'= 6$, as
$t'/t$ varies, setting the the energy scale $w=6$ throughout the paper.
As illustrated in Fig.~\ref{fig4}(a), the band gap $\Delta$
vanishes for $t'/t<2$.  These bands touch at two
Dirac points for $t'/t = \frac{1}{2}$ in
Fig.~\ref{fig4}(b).
Figure~\ref{fig4}(c)
shows the band insulating case, $t/t'=0.25$.

To characterize the magnetic properties of Eq.~\ref{hubham} we
measure the AF structure factor
\begin{align}
S_{\rm AF} =
\frac{1}{N} \sum_{l,j} (-1)^{l}
\langle \vec S_{j} \cdot \vec S_{l+j} \rangle
\end{align}
where the factor $(-1)^{l} = +1(-1)$ if site $l$ is
on the same(different)
sublattice of the bipartite structure of
Fig.~\ref{fig1}.

The spin correlation in the singlet phase
falls off exponentially with separation $l$ and $S_{\rm AF}$
is independent of lattice size.
If LRAFO is present, $S_{\rm AF} \propto N$,
since spin correlations remain nonzero out to all distances on
a finite lattice.

\begin{figure}[t]
   \includegraphics[width=7.0cm]{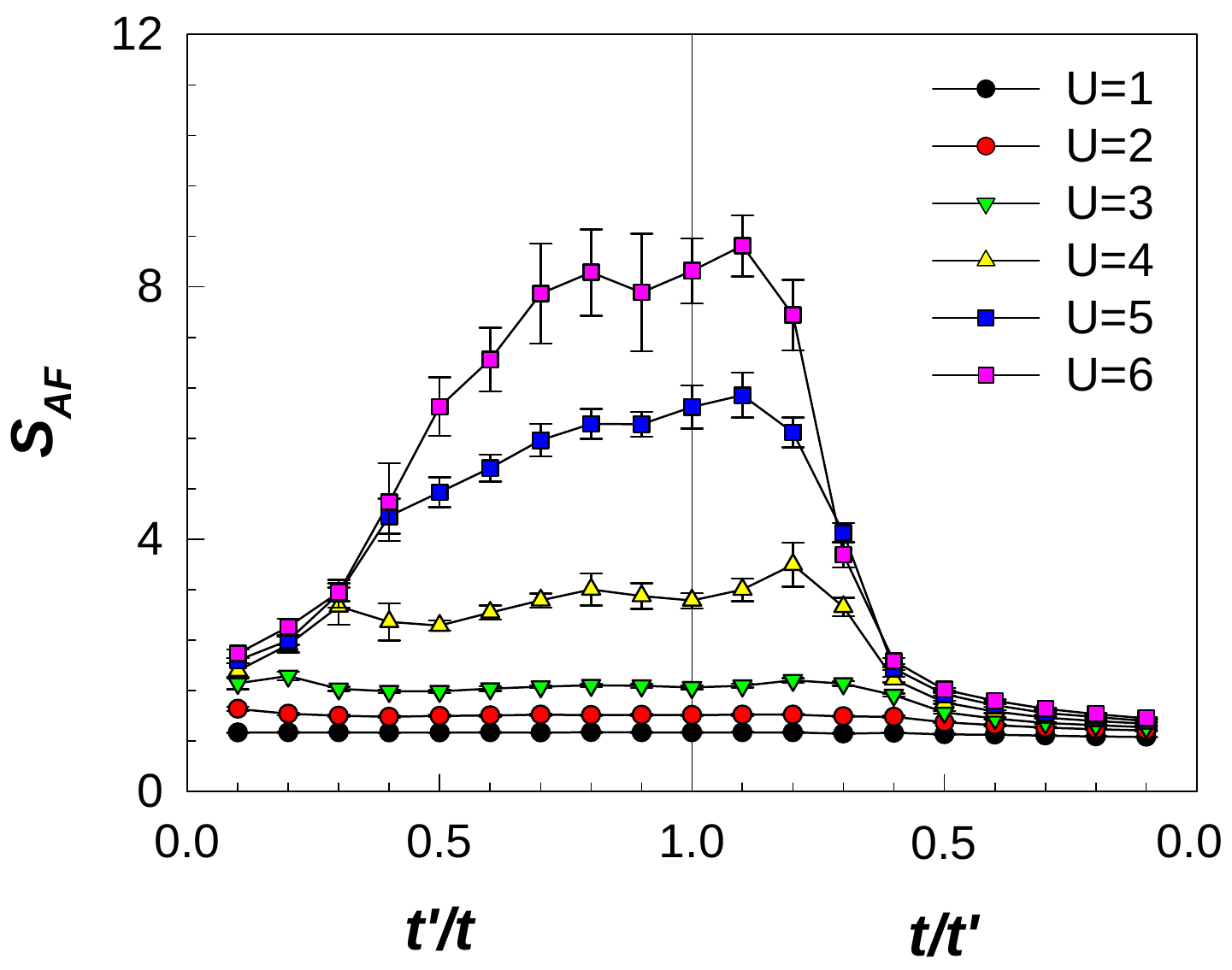}
   \caption{The AF structure factor $S_{\rm AF}$
is shown as a function of hopping anisotropy for different $U$.
The linear lattice size $L=8$ so that the number of
sites $N=128$.
(There are 64 unit cells each with two sites).
The inverse temperature discretization $\Delta \tau = \beta/L = 1/2U$ except for $U=1$ where $\Delta \tau=1/4$.
Data were acquired from 25 simulations of 1000 equilibration and
4000 measurement sweeps for each $t'/t$.
\label{fig5}}
   \end{figure}

Figure \ref{fig5} shows $S_{\rm AF}$ on an $N=8\times 8$ lattice
for different $U$ as a function of $t/t'$.
It is known that LRAFO exists at the symmetric honeycomb
lattice point $t=t'$ only when $U$ is sufficiently large
\cite{paiva05,meng10,hohenadler12,zheng11}, with the
most accurate value\cite{sorella12}
of the critical point $U_c = 3.869 \pm 0.013$.
The data of Fig.~\ref{fig5} is suggestive of this
result, with $S_{\rm AF}$ being essentially
independent of the value of $t/t'$ for $U=1,2,3$,
and becoming both larger and sensitive to the anisotropy
for $U \geq 4$.

\begin{figure}[!h]
   \includegraphics[width=8.0cm]{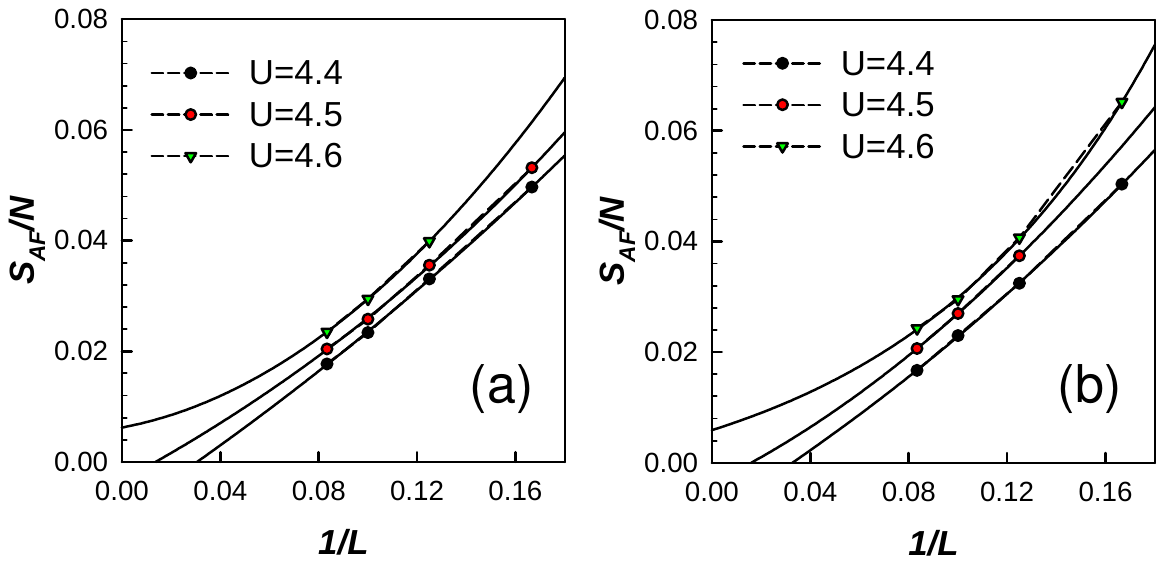}
   \caption{Finite size scaling of the AF structure factor $S_{\rm AF}$
for $t/t'=0.95$ (a) and $t'/t$=0.90 (b).
In both cases $U_c > 4.5$ is well above the critical interaction
strength $U_c=3.869$ for isotropic hopping\cite{sorella12}.
\label{fig6}}
   \end{figure}

Finite size scaling can be used to analyze quantitatively the
possibility of LRAFO.  Such data are shown in Fig.~\ref{fig6}.
We find that hopping anisotropy increases $U_c$,
in agreement with our results for the $g$ dependence of the
order parameter in the strong coupling Heisenberg model
(Fig.~\ref{fig3}) which falls off to either side of $g=1$.

\begin{figure}[!h]
   \includegraphics[width=8.5cm]{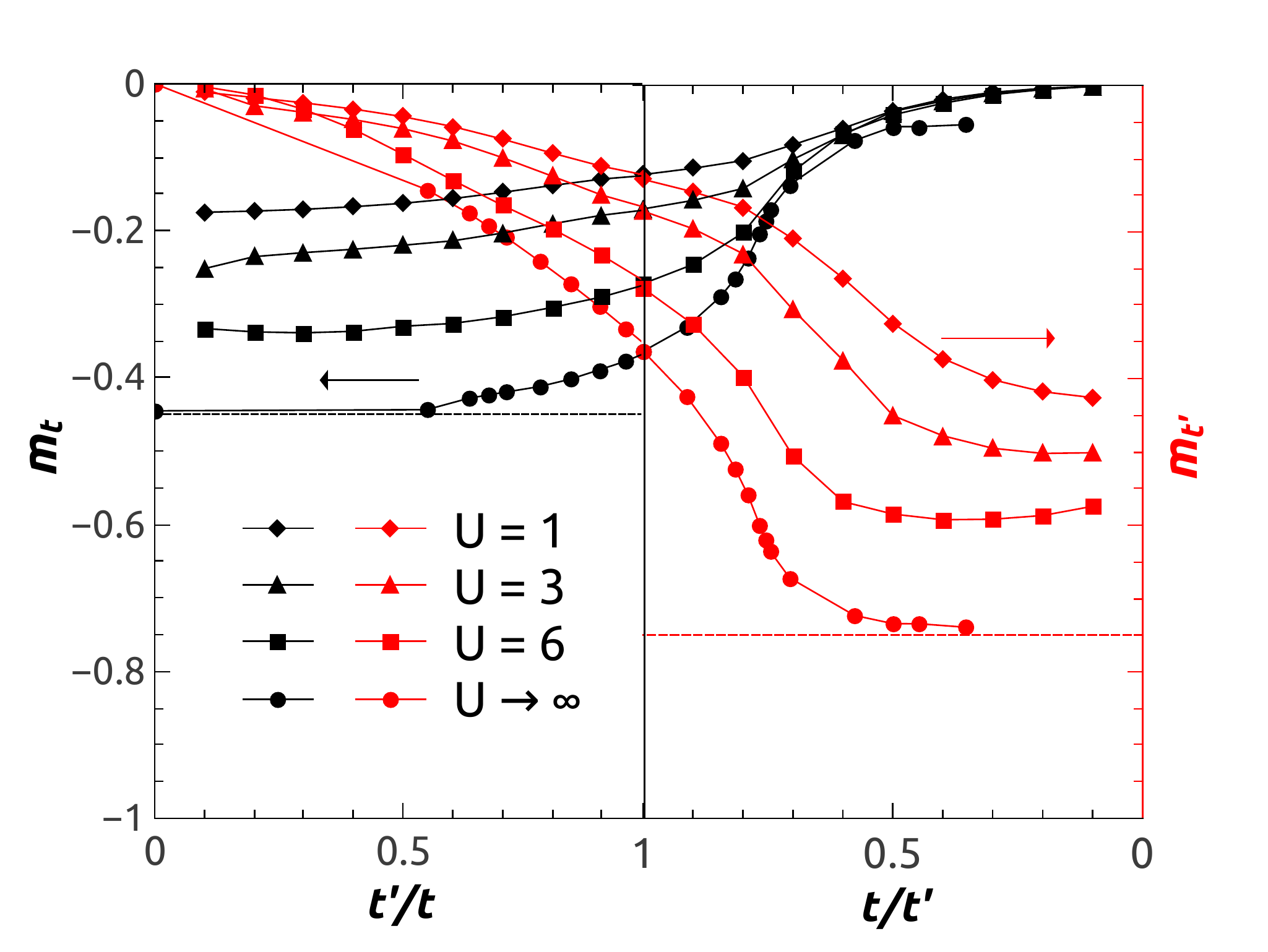}
   \caption{Near neighbor (singlet) spin correlation function
across intra- and inter-chain bonds, $m_t$ and $m_{t'}$, respectively.
$\langle S_i \cdot S_j \rangle$ is large and independent of
$t'/t$ for $t'/t \gtrsim 2$.
This value matches the point at which a nonzero gap $\Delta$
opens in the spectrum, Fig.~\ref{fig4}(a). The limiting value at $t'=0$ ($t=0$) is $0.4515$\cite{fisher} ($0.75$).
   \label{fig7}
}
   \end{figure}

A second diagnostic of magnetic order is the near-neighbor spin
correlation between adjacent pairs of sites.
This can be evaluated for both intra- and inter-chain bonds,
and measures the formation of singlet correlations,
$m_t$ and $m_{t'}$ respectively,
on the associated bonds.
Fig.~\ref{fig7} shows $m_t$  and $m_{t'}$  for different values of $U$.
For the Heisenberg limit, $U \to \infty$,
we use $J \sim t^2/U$ to convert $g =J'/J$ to $\sqrt{t'/t}$.
In the strong coupling limit
$\langle S_i \cdot S_j \rangle = -\frac{3}{4}$ for
a singlet.  Here in the Hubbard model, the finite
value of the on-site repulsion,  $U < \infty$, allows
for charge fluctuations which reduce the magnitude of the
singlet correlator.
The quantities $m_t$ and $m_{t'}$ have opposite trends
in the two regimes $t'<t$ and $t<t'$ of Fig.~\ref{fig7}.
When $t/t' < 1$, $m_t$ is suppressed, and $m_{t'}$ increases and
saturates with decreasing $t/t'$.
This supports the physical scenario in which singlets are
formed between the stronger $t'$ bonds.
On the other hand, if $t'/t<1$, $m_{t'}$ is diminshed.
$m_t$ approaches the short range AF correlations of the 1-d chains \cite{fisher},
without the formation of singlets on the $t$ bonds.
Thus although at first glance Fig.~\ref{fig5} indicates similar, reduced
values for
$S_{\rm AF}$ for both small $t'/t$ and for small $t/t'$,
the singlet correlator of Fig.~\ref{fig7} suggests these
are rather distinct limits:  full singlets form at
$t/t' \rightarrow 0$ but not
$t'/t \rightarrow 0$.

\begin{figure}[!h]
   \includegraphics[width=7.0cm]{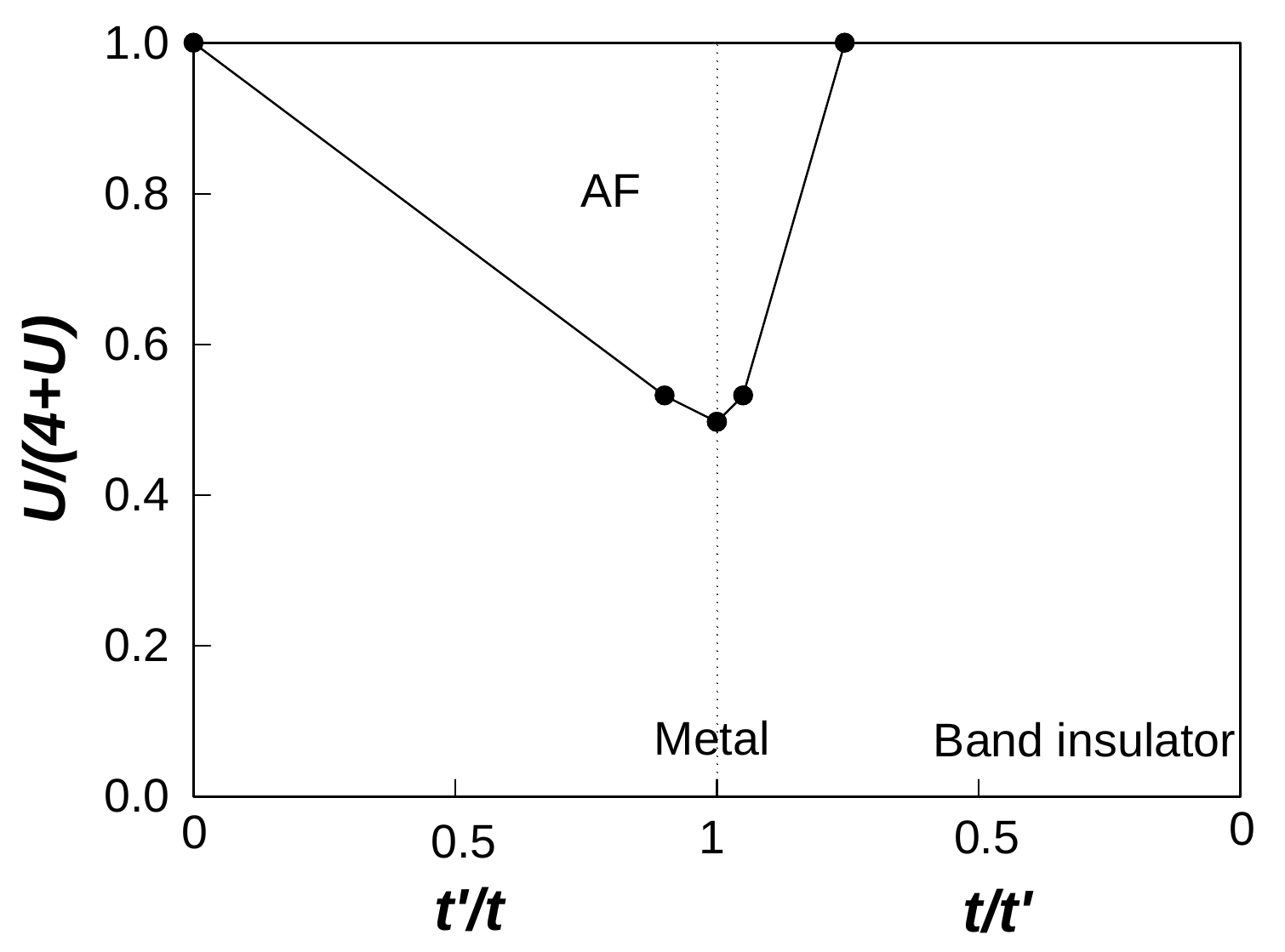}
   \caption{Phase diagram.  The $U=\infty$ Heisenberg limit is
along the top of the figure, $U/(4+U)=1$, and is extracted from the
data of Fig.~\ref{fig3}.
The critical interaction strength diverges even prior to
entry into the band insulator phase at $t/t'=0.5$.
 \label{fig8}
}
   \end{figure}

The evaluation of these magnetic correlations allows us to sketch the
phase diagram in the plane of hopping anisotropy and interaction
strength shown in Fig.~\ref{fig8}.
The fact that $g_c = 1.75$ in the Heisenberg limit is less than
the anisotropy required to open a nonzero gap $\Delta$ in the
non-interacting band structure suggests that the destruction of
LRAFO involves more than the simple RPA-like criterion of the
vanishing of the density of states at the Fermi level.  That is,
the competing possibility of singlet formation also plays a role in
the absence of LRAFO.

\section{Conclusion}

In this paper we have investigated magnetic ordering on a two
dimensional lattice formed by the regular removal of one third of
the sites from a square lattice.  We analyzed the strong coupling,
Heisenberg limit using spin-wave theory and QMC (SSE),
and determined the range of the ratio $J'/J$ on the two types of bonds
in which an ordered AF phase exists at $T=0$.  Unlike the
one fifth depleted lattice, which breaks into small clusters
in both the $J=0$ and $J'=0$ limits, we have shown that
AF order persists to very small $J'/J$
as a consequence of the fact that extended one dimensional chains
are still present when $J'=0$.

We also used DQMC to study the single band Hubbard Hamiltonian
on this lattice.  The singlet correlator was found to grow
rapidly for $t'/t \sim 1.5$, coinciding with a loss of AF order and
the approach to the band insulator at $t'/t>2$ in the noninteracting limit.
The critical interaction strength $U_c \sim 3.87$ for $t=t'$
was shown to increase with inhomogeneity $t' \neq t$.
The effect of {\it random} removal of sites on AF order has been
studied in both itinerant and localized models
\cite{ulmke99,hoglund04,sigrist96,hass01,mendels09}.

The one third depleted geometry that we investigated
has recently been
shown to be realized as a result of charge stripe ordering
in the nickelates\cite{zhang16,botana16}, so our simulations
speak to the conditions for AF order in those materials.
The relative strengths of first and second neighbor exchange couplings
for nickelates has not yet been addressed.
Another key feature is the presence of multiple NiO$_2$ layers
and the surprising nature of charge equivalence between the
layers\cite{zhang16,botana16}.
We cannot immediately address this phenomenon, since
in our treatment charge ordering is put in {\it a priori}
through our consideration of a one third depleted lattice and,
in addition, our restriction to a single layer model.

A more approximate method than DQMC, which considers
itinerant electrons interacting with classical spins\cite{sen06,sen10}
can be employed to treat multiple bands. It may be used to explore
the {\it spontaneous} formation of charge ordering, and we leave the details of this to future study.

\vskip0.15in
\section{Acknowledgements:}
H.G. acknowledges support from China Scholarship Council. T.M.~acknowledges
funding from Science Without Borders, Brazil.  The work of W.E.P.~was supported by DOE grant DE-FG02-04ER46111.
The work of R.T.S.~was supported by DOE grant DE-SC0014671.

\renewcommand{\thefigure}{S\arabic{figure}}
\setcounter{figure}{0}


\begin{thebibliography}{100}

\bibitem{white89a}
``A Numerical Study of the Two--Dimensional Hubbard Model
with Repulsive Coulomb Interaction,"
S.R.~White, D.J.~Scalapino, R.L.~Sugar,
E.Y.~Loh, Jr., J.E.~Gubernatis, and R.T.~Scalettar,
Phys.~Rev.~ B40, 506 (1989).

\bibitem{scalapino94}
D.J. Scalapino, Does the Hubbard Model Have the Right Stuff?  in
``Proceedings of the International School of Physics", edited
by R.A. Broglia and J.R. Schrieffer (North-Holland, New York, 1994),
and references cited therein.

\bibitem{vilk97}
``Non-Perturbative Many-Body Approach to the Hubbard Model and
Single-Particle Pseudogap",
Y.M. Vilk and A.-M.S. Tremblay,
J. Phys. I 7, 1309 (1997).

\bibitem{zhang97b}
``Pairing Correlations in the Two-Dimensional Hubbard Model,"
S. Zhang, J. Carlson, J.E. Gubernatis,
Phys. Rev. Lett. 78, 4486 (1997).

\bibitem{maier00}
``$d$-Wave Superconductivity in the Hubbard Model", Th. Maier, M.
Jarrell,
Th. Pruschke, and J. Keller, Phys. Rev. Lett. 85, 1524 (2000).

\bibitem{capone07}
``Cellular-dynamical mean-field theory of the competition between
antiferromagnetism and d-wave superconductivity in the two-dimensional
Hubbard model", M. Capone and G. Kotliar, J. Mag. and
Mag. Mat. 310, 529 (2007).

\bibitem{gull07}
``Momentum space anisotropy and pseudogaps: a comparative cluster
dynamical mean field analysis of the doping-driven metal-insulator
transition in the two dimensional Hubbard model,"
E.~Gull, M.~Ferrero, O.~Parcollet, A.~Georges, and
A.J.~Millis,
Phys.~Rev.~B82, 155101 (2007).

\bibitem{berg09}
``Striped superconductors: how spin, charge and superconducting orders
intertwine in the cuprates,"
E. Berg, E. Fradkin, S.A. Kivelson and J.M. Tranquada,
New J. Phys. 11, 115004 (2009).

\bibitem{lieb89}
``Two Theorems on the Hubbard Model,"
E.H. Lieb,
Phys. Rev. Lett. 62, 1201 (1989).

\bibitem{n_katoh_95}
``Spin Gap in 2-Dimensional Heisenberg-Model for CaV$_4$O$_9$,"
N. Katoh and M. Imada,
J.  Phys. Soc. Jpn. 64, 4105 (1995).

\bibitem{k_ueda_96}
``Plaquette Resonating-Valence-Bond Ground State of CaV$_4$O$_9$,"
K.~Ueda, H.~Kontani, M.~Sigrist, and P.~A.Lee,
Phys. Rev. Lett. 76, 1932 (1996).

\bibitem{m_troyer_96}
``Phase Diagram of Depleted Heisenberg Model for CaV$_4$O$_9$,"
M.~Troyer, H.~Kontani, and K.~Ueda,
Phys. Rev. Lett. 76, 3822 (1996).

\bibitem{m_gelfand_96}
``Convergent Expansions for Properties of the Heisenberg Model for
CaV$_4$O$_9$,"
M.~P. Gelfand, Weihong Zheng, R.R.P.~Singh, J.~Oitmaa,
and C.~J. Hamer,
 Phys. Rev. Lett. 77, 2794 (1996).

\bibitem{cavo_wep}
``Impact of Structure on Magnetic Coupling in CaV$_4$O$_9$,"
W.~E. Pickett,
Phys. Rev. Lett. 79, 1746 (1997).

\bibitem{w_bao_11}
``A Novel Large Moment Antiferromagnetic Order in
K$_{0.8}$Fe$_{1.6}$Se$_{2}$
Superconductor,"
B.~Wei, H.~Qing-Zhen, C.~Gen-Fu, M.~A. Green, W.~Du-Ming, H. Jun-Bao,
and Q.~Yi-Ming,
Chin. Phys. Lett. 28, 086104 (2011).

\bibitem{f_ye_11}
``Common Crystalline and Magnetic Structure of Superconducting
A$_2$Fe$_4$Se$_5$
(A=K,Rb,Cs,Tl) Single Crystals Measured Using Neutron Diffraction,"
F.~Ye, S.~Chi, W.~Bao, X.~F.~Wang, J.~J.~Ying, X.~H.~Chen, H.~D.~Wang,
C.~H.~Dong, and M.~Fang,
Phys. Rev. Lett. 107, 137003 (2011).

\bibitem{park09}
``Making massless Dirac fermions from a
patterned two-dimensional electron gas,"
C.H. Park and S.G. Louie,
Nano Lett. 9, 1793 (2009).

\bibitem{gomes12}
``Designer Dirac fermions and topological phases in molecular graphene,"
K.K. Gomes, W. Mar, W. Ko, F. Guinea, and H.C. Manoharan,
Nature 483, 306 (2012).

\bibitem{polini13}
``Artificial honeycomb lattices for electrons, atoms and photons,"
M. Polini, F. Guinea, M. Lewenstein, H.C. Manoharan, and V. Pellegrini,
Nature Nanotechnology 8, 625 (2013).

\bibitem{hou07}
``Electron fractionalization in two-dimensional graphene-like
structures,"
C.-Y. Hou, C. Chamon, and C. Mudry,
Phys. Rev. Lett. 98, 186809 (2007).

\bibitem{roy10}
``Unconventional superconductivity on honeycomb
lattice: theory of Kekul\'e order parameter,"
B. Roy and I.F. Herbut,
Phys. Rev. B82, 035429 (2010).

\bibitem{zhang16}
``Stacked charge stripes in the quasi-2D trilayer nickelate
La$_4$Ni$_3$O$_8$,"
J. Zhang, Y.-S. Chen, D. Phelan, H. Zheng, M.R, Norman, and J.F.
Mitchell,
Proc. Nat. Acad. Sci. 113, 8945 (2016).

\bibitem{botana16}
``Charge ordering in Ni$^{1+}$/Ni$^{2+}$ nickelates:
La$_4$Ni$_3$O$_8$,"
and La$_3$Ni$_2$O$_6$,"
A.S. Botana, V. Pardo, W.E. Pickett and M.R. Norman,
Phys. Rev. B94, 081105(R) (2016).

\bibitem{tranquada94}
``Simultaneous Ordering of Holes and Spins in La$_2$NiO$_{4.125}$,"
J.M. Tranquada, D.J. Buttrey, V. Sachan, and J.E. Lorenzo,
Phys. Rev. Lett. 73, 1003 (1994).

\bibitem{sachan95}
``Charge and spin ordering in La$_{2-x}$Sr$_x$NiO$_{4.00}$ with $x=0.135$
and 0.20,"
V. Sachan, D.J. Buttrey, J.M. Tranquada, J.E. Lorenzo, and G. Shirane,
Phys. Rev. B51, 12742 (1995).

\bibitem{yoshizawa00}
``Stripe order at low temperatures in La$_{2-x}$Sr$_x$NiO$_4$ with
$0.289 \leq x \leq 0.5$,"
H. Yoshizawa, T. Kakeshita, R. Kajimoto, T. Tanabe, T. Katsufuji, and Y.
Tokura,
Phys. Rev. B61, R854(R) (2000).

\bibitem{kajimoto03}
``Spontaneous rearrangement of the checkerboard charge order to stripe
order in La$_{1.5}$Sr$_{0.5}$NiO$_4$,"
R. Kajimoto, K. Ishizaka, H. Yoshizawa, and Y. Tokura,
Phys. Rev. B67, 014511 (2003).

\bibitem{hotta04}
``Orbital Ordering, New Phases, and Stripe Formation in Doped Layered
Nickelates,"
T. Hotta and E. Dagotto,
Phys. Rev. Lett. 92, 227201 (2004).

\bibitem{syljuasen02}
``Quantum Monte Carlo with Directed
Loops", O.F. Syljuasen and A. W. Sandvik,
Phys. Rev. E66, 046701 (2002).

\bibitem{sandvik94}
``Order-disorder transition in a two-layer quantum antiferromagnet,"
A.W. Sandvik and D.J. Scalapino,
Phys. Rev. Lett. 72, 2777 (1994).



\bibitem{affleck94}
``A plane of weakly coupled Heisenberg chains: theoretical arguments and
numerical calculations,"
I. Affleck, M.P. Gelfand, and R.R.P. Singh,
J. Phys. Math. Gen. 27, 7313 (1994).

\bibitem{Sakai89}
``The Ground State of Quasi-One-Dimensional Heisenberg Antiferromagnets,"
T. Sakai and M. Takahashi, J. Phys. Soc. Japan 58, 3131 (1989).

\bibitem{sandvik99}
``Multichain Mean-Field Theory of Quasi-One-Dimensional Quantum Spin
Systems,"
A.W. Sandvik,
Phys. Rev. Lett. 83, 3069 (1999).

\bibitem{blankenbecler81}
``Monte Carlo calculations of coupled boson-fermion systems. I,"
R. Blankenbecler, R. L. Sugar, and D. J. Scalapino,
Phys. Rev. D24, 2278 (1981).

\bibitem{loh90}
``The Sign Problem in the Numerical Simulation of Many
Electron Systems,"
E.Y.~Loh, J.E.~Gubernatis, R.T.~Scalettar, S.R.~White,
D.J.~Scalapino, and R.L.~Sugar,
Phys.~Rev.~B41, 9301 (1990).

\bibitem{paiva05}
``Ground-state and finite-temperature signatures of quantum phase
transitions in the half-filled Hubbard model on a honeycomb lattice,"
T. Paiva, R.T. Scalettar, W. Zheng, R.R.P. Singh, and J. Oitmaa,
Phys. Rev. B72, 085123 (2005).

\bibitem{meng10}
``Quantum spin liquid emerging in two-dimensional correlated Dirac
fermions,"
Z. Meng, T. Lang, S. Wessel, F. Assaad, and A. Muramatsu,
Nature (London) 464, 847 (2010).

\bibitem{hohenadler12}
``Quantum phase transitions in the Kane-Mele-Hubbard model,"
M. Hohenadler, Z.Y. Meng, T.C. Lang, S. Wessel, A.
Muramatsu, and F.F. Assaad, Phys. Rev. B85, 115132 (2012).

\bibitem{zheng11}
``Particle-hole symmetry and interaction effects in the Kane-Mele-Hubbard
model,"
D. Zheng, G.-M. Zhang, and C. Wu, Phys. Rev. B84, 205121
(2011).

\bibitem{sorella12}
``Absence of a Spin Liquid Phase in the Hubbard Model on the Honeycomb
Lattice,"
S. Sorella, Y. Otsuka, and S. Yunoki, Sci. Rep. 2, 992 (2012).

\bibitem{fisher}
``Linear magnetic chains with anisotropic coupling," J.C. Bonner and
M.E. Fisher,
Phys. Rev. 135, A640 (1964).

\bibitem{ulmke99}
``Disorder and Impurities in Hubbard Antiferromagnets",
M.~Ulmke, P.~J.~H.~Denteneer, V.~Janis, R.~T.~Scalettar, A.~Singh,
D.~Vollhardt, and G.~T.~Zimanyi,
Adv. Sol. St. Phys. 38, 369 (1999).

\bibitem{hoglund04}
``Impurity effects at finite temperature in the two-dimensional
$S=1/2$ Heisenberg antiferromagnet,"
K.H. Hoglund and A.W. Sandvik,
Phys. Rev. B70, 024406 (2004).

\bibitem{sigrist96}
``Low-Temperature Properties of the Randomly Depleted
Heisenberg Ladder'',
M. Sigrist and A. Furusaki,
J. Phys. Soc. Jpn. 65, 2385 (1996)

\bibitem{hass01}
``Order by Disorder from Nonmagnetic Impurities in
a Two-Dimensional Quantum Spin Liquid'',
S. Wessel, B. Normand, M. Sigrist, and S. Haas,
Phys. Rev. Lett. 86, 1086 (2001).

\bibitem{mendels09}
``Impurity-Induced Magnetic Order in Low-Dimensional
Spin-Gapped Materials'',
J. Bobroff, N. Laflorencie, L. K. Alexander, A.V. Mahajan, B.
Koteswararao, and P.  Mendels, Phys. Rev. Lett. 103, 047201 (2009).

\bibitem{sen06}
``Colossal magnetoresistance observed in Monte Carlo simulations of the
one- and two-orbital models for manganites," C. Sen, G. Alvarez, H.
Aliaga, and E. Dagotto, Phys. Rev. B73, 224441 (2006).

\bibitem{sen10}
First Order Colossal Magnetoresistance Transitions in the Two-Orbital
Model for Manganites",
C. Sen, G. Alvarez, and E. Dagotto,
Phys. Rev. Lett. 105, 097203 (2010).


\end{thebibliography}
\end{document}